\documentclass[longbibliography,
 reprint,
superscriptaddress,
 amsmath,amssymb,
 aps,
]{revtex4-2}

\usepackage{graphicx}
\usepackage{dcolumn}
\usepackage{bm}
\usepackage{multirow}

\newcommand{\EE}[2]{\ensuremath{{#1}\times 10^{#2}}}

\renewcommand{\apj}{The Astrophysical Journal}
\newcommand{\npa}{Nuclear Physics A}
\newcommand{\nima}{Nuclear Instruments and Methods A}

\begin{document}


\title{$\beta$-decay Measurements Near the $N=40$ Island of Inversion to Quantify Cooling of Accreted Neutron Star Crusts}

\author{K. Hermansen}
\affiliation{Department of Physics and Astronomy, Michigan State University, East Lansing, MI 48823, USA}
\affiliation{Facility for Rare Isotope Beams, Michigan State University, East Lansing, MI 48823, USA}

\author{W.-J. Ong}
\affiliation{Nuclear and Chemical Sciences Division, Lawrence Livermore National Laboratory, Livermore, California 94550, USA}

\author{H. Schatz}
 \email{schatz@msu.edu}
\affiliation{Department of Physics and Astronomy, Michigan State University, East Lansing, MI 48823, USA}
\affiliation{Facility for Rare Isotope Beams, Michigan State University, East Lansing, MI 48823, USA}


\author{J. Browne}
\affiliation{Department of Physics and Astronomy, Michigan State University, East Lansing, MI 48823, USA}
\affiliation{Facility for Rare Isotope Beams, Michigan State University, East Lansing, MI 48823, USA}

\author{A. Chester}
\affiliation{Facility for Rare Isotope Beams, Michigan State University, East Lansing, MI 48823, USA}

\author{K. Childers}
\affiliation{Department of Physics and Astronomy, Michigan State University, East Lansing, MI 48823, USA}
\affiliation{Facility for Rare Isotope Beams, Michigan State University, East Lansing, MI 48823, USA}


\author{R. Jain}
\affiliation{Department of Physics and Astronomy, Michigan State University, East Lansing, MI 48823, USA}
\affiliation{Facility for Rare Isotope Beams, Michigan State University, East Lansing, MI 48823, USA}


\author{S. Liddick}
\affiliation{Department of Chemistry, Michigan State University, East Lansing, MI 48823, USA}
\affiliation{Facility for Rare Isotope Beams, Michigan State University, East Lansing, MI 48823, USA}

\author{S. Lyons}
\altaffiliation[Current address: ]{Pacific Northwest National Laboratory, Richland WA 99354, USA}
\affiliation{Facility for Rare Isotope Beams, Michigan State University, East Lansing, MI 48823, USA}

\author{S. A. Miskovich}
\altaffiliation[Current address: ]{SLAC National Accelerator Laboratory, Menlo Park, CA, USA}
\affiliation{Department of Physics and Astronomy, Michigan State University, East Lansing, MI 48823, USA}
\affiliation{Facility for Rare Isotope Beams, Michigan State University, East Lansing, MI 48823, USA}

\author{P. M{\"o}ller}
\affiliation{Mathematical Physics, Lund University, S-221 00 Lund, Sweden \\}

\author{F. Montes}
\affiliation{Facility for Rare Isotope Beams, Michigan State University, East Lansing, MI 48823, USA}

\author{J. Owens-Fryar}
\affiliation{Department of Physics and Astronomy, Michigan State University, East Lansing, MI 48823, USA}
\affiliation{Facility for Rare Isotope Beams, Michigan State University, East Lansing, MI 48823, USA}

\author{A. Palmisano-Kyle}
\affiliation{Department of Nuclear Engineering, University of Tennessee, Knoxville, TN 37996, USA}

\author{A. L. Richard}
\affiliation{Department of Physics and Astronomy, Michigan State University, East Lansing, MI 48823, USA}
\affiliation{Facility for Rare Isotope Beams, Michigan State University, East Lansing, MI 48823, USA}
\affiliation{Department of Physics and Astronomy, Ohio University, Athens, OH 45701, USA}

\author{N. Rijal}
\affiliation{Facility for Rare Isotope Beams, Michigan State University, East Lansing, MI 48823, USA}
\affiliation{Nuclear Center of Excellence, Huston Formation Evaluation, SLB, Sugarland, Texas 77478, USA}

\author{M. Smith}
\affiliation{Facility for Rare Isotope Beams, Michigan State University, East Lansing, MI 48823, USA}

\author{D. Soltesz}
\affiliation{Department of Physics and Astronomy, Ohio University, Athens, OH 45701, USA}

\author{A. Spyrou}
\affiliation{Department of Physics and Astronomy, Michigan State University, East Lansing, MI 48823, USA}
\affiliation{Facility for Rare Isotope Beams, Michigan State University, East Lansing, MI 48823, USA}

\author{S. K. Subedi}
\affiliation{Department of Physics and Astronomy, Ohio University, Athens, OH 45701, USA}

\author{L. Wagner}
\affiliation{Facility for Rare Isotope Beams, Michigan State University, East Lansing, MI 48823, USA}


\date{\today}

\begin{abstract}
Understanding the thermal structure of the outer crust of accreting neutron stars is important to interpret astronomical X-ray observations. Ground-state to ground-state $\beta$-decay transitions of neutron-rich nuclei comprising the crust enable Urca neutrino cooling processes that affect this thermal structure. Here we constrain the ground-state to ground-state transition strengths for the decays of $^{57}$Sc, $^{57}$Ti, and $^{59}$Ti based on experimental data. The data were obtained by combining total absorption $\gamma$-spectroscopy data from the SuN detection system with $\beta$-delayed neutron emission data from the NERO detection system at Michigan State University's National Superconducting Cyclotron Laboratory. We find $\log ft=$5.8$^{+0.3}_{-0.2}$ and $\log ft=$5.34$^{+0.08}_{-0.24}$ for the decays of $^{57}$Ti and $^{59}$Ti, respectively, and find no evidence for ground-state feeding in the decay of $^{57}$Sc. The results indicate weaker transitions than predicted by theory and indicated by previous measurements, resulting in reduced efficiency of neutrino cooling in accreted neutron star crusts in systems that exhibit X-ray superbursts.

\end{abstract}

\maketitle


\section{\label{sec:intro}Introduction}

Neutron stars in stellar binary systems can accrete matter from the companion star, resulting in bright X-ray emission from the neutron star surface. Quasi-persistent soft X-ray sources are a subset of systems where extended periods of accretion lasting years to decades alternate with extended periods of quiescence, where accretion stops and X-ray emission is reduced by many orders of magnitude \cite{Meisel2018,Wijnands2017}. During accretion, nuclear reactions throughout the crust of the neutron star heat the crust to typical temperatures of 10$^8$ to 10$^9$K. During quiescence, X-ray observations can track the cooling of the heated crust over timescales of months to years. The details of these cooling curves probe the properties of dense matter comprising the outer layers of the neutron star such as the presence of superfluid neutrons \citep{Shternin2007Neutron-star-co,Brown2009Mapping-Crustal}, the structure of the crystalline lattice forming the crust \citep{Cackett2006Cooling-of-the-,Shternin2007Neutron-star-co,Brown2009Mapping-Crustal}, or the existence of a disordered nuclear pasta layer \citep{Horowitz2015,Deibel2017}.

To interpret observations through model comparisons, the nuclear reactions that heat and cool the crust during accretion and during quiescence need to be understood. Here we present experimental results related to the strength of nuclear Urca processes in accreted neutron star crusts that can lead to strong neutrino cooling \cite{Schatz2013, Deibel2016}. Owing to their steep $T^5$ temperature dependence, crust Urca processes limit the maximum crust temperature for a given heating rate and enhance early cooling. Crust Urca cooling has been shown to affect the interpretation of observed cooling curves, especially in very hot systems such as MAXI J0556-332 \cite{meiselConstraintsBygoneNucleosynthesis2017}. The Urca processes occur at composition interfaces where the electron Fermi energy is within $\approx kT$ of the electron-capture threshold, and the composition switches to the more neutron-rich electron-capture daughter with increasing depth. At the interface a thin layer can form where both, electron capture and $\beta$ decay are possible, leading to rapid cooling via neutrino emission from alternating electron capture and $\beta$-decay cycles. Such Urca cycles can only occur when the $\beta$-decay transitions proceed to the daughter ground-state (gs) or to very low-lying excited states with energies of the order of $kT$ (typically 10--100 keV, $T$ being the ambient temperature in the neutron star). Only for those transitions can electrons acquire the high energies needed to fill the small open phase space near the Fermi energy. The strength of Urca cooling for a particular pair of parent and daughter nuclei is therefore determined by the gs-gs $\beta$-decay transition strengths. 

Which nuclear species are present at the relevant depths depends mostly on the initial mass number range produced by hydrogen, helium, and carbon burning on the neutron star surface, for example via X-ray bursts. Ongoing accretion will increase the density and induce electron captures that transform the nuclear ashes mostly along isotopic chains with constant mass number. At greater depth neutron transfer among nuclei can alter the mass chain pattern somewhat \cite{Schatz2022}. Odd-A mass chains tend to be much more important, as even-A mass chains are characterized by 2-step electron captures due to the odd-even staggering of the energy thresholds. As a consequence, Urca cooling is strongly suppressed: at the first electron-capture transition because the second transition is typically faster than the $\beta$ decay, and at the second transition because the $\beta$ decay is Pauli blocked owing to the much lower Q-value. 

Crust models rely mostly on gs-gs electron capture and $\beta$-decay transition strength predictions from the QRPA-fY model \cite{mollerNewDevelopmentsCalculation1990,mollerNuclearPropertiesAstrophysical1997, Moller2019}. This model provides predictions across the entire nuclear chart. Recently, \citet{ongDecay61VIts2020} determined gs-gs log ft transition for Urca cooling via $^{61}$Cr$\leftrightarrow ^{61}$V, which is important for systems with mixed H/He bursts where the rp-process produces a broad range of nuclei, including with $A=61$. They found that QRPA-fY strength predictions for  $^{61}$V $\beta$ decay significantly overestimated the gs-gs transition strength. Here we focus on slightly lower masses of particular importance for accreting neutron star systems that exhibit superbursts. Superbursts are thought to be powered by deep explosive burning of $^{12}$C and produce nuclear ashes with masses in the A=52--59 range. 

Concerning the important odd-A mass chains, for $A=53,55$ currently recommended ground-state spins constrained by experimental data, and theoretical predictions at greater depth indicate no significant allowed gs-gs transitions. 

For $A=57$, there are three potential strong Urca pairs: $^{57} \rm{Cr} \leftrightarrow ^{57}$V,
$^{57} \rm{V} \leftrightarrow ^{57}$Ti, and  $^{57}\rm{Ti} \leftrightarrow ^{57}$Sc. For $^{57}$Cr$ \leftrightarrow ^{57}$V, high-resolution $\gamma$-spectroscopy experiments following the $\beta$ decay of $^{57}$V indicate a large gs-gs branch of 21\% \cite{mantica$ensuremathbeta$DecayStudies2003} resulting in a significant log$ft$ transition strength of 5.1, albeit weaker than the 4.5 predicted by QRPA-fY theory. For $^{57}$V $\leftrightarrow ^{57}$Ti, high-resolution $\gamma$-spectroscopy following the $\beta$ decay of $^{57}$Ti indicates a very large gs-gs branch of 54\% \cite{liddickVdecayOddA57Ti2005} resulting in a fast log$ft$ transition strength of 4.7, faster than the 5.3 predicted by QRPA-fY, though QRPA-fY does predict a log$ft$=4.7 transition to an excited state at 80~KeV, which may be the ground state within theoretical uncertainties. However, the $\beta$-decay Q-value is with 10~MeV relatively large, and the high-resolution $\gamma$-spectroscopy measurements may suffer from the so-called Pandemonium effect \cite{hardyEssentialDecayPandemonium1977}. This effect occurs when a large number of weak feeding transitions that are missed in the
high-resolution spectra add up to substantial feeding of specific excited states, thus leading to a potential overestimation of the ground-state transition strength. For $^{57}$Ti $\leftrightarrow ^{57}$Sc, five $\gamma$-transitions have been identified in $^{57}$Ti following the decay of $^{57}$Sc \cite{crawfordDecayIsomericProperties2010b} but statistics was insufficient to deduce a level scheme or $\beta$-feeding branchings. Level systematics indicate possible ground-state spins of 5/2$^-$ and 7/2$^-$  for $^{57}$Ti and $^{57}$Sc, respectively, and thus the possibility of a strong allowed gs-gs transition. 

In the $A=59$ electron-capture chain there are also several potential strong Urca pairs. $\beta$-delayed high-resolution $\gamma$-spectroscopy studies of the decay of $^{59}$Mn indicate a significant gs-gs branch of 23\% and an associated log$ft$=5.44 \cite{oinonenGroundstateSpin59Mn2001} making $^{59}$Fe$\leftrightarrow ^{59}$Mn an Urca pair. In addition, based on current tentative spin assignments for $^{59}$V (5/2$^-$,3/2$^-$) and $^{59}$Ti (5/2$^-$) \cite{BASUNIA20181}, $^{59}$V$\leftrightarrow ^{59}$Ti are also a potential strong Urca pair.

We present here new data for $\beta$-decay transition strengths for the decays of $^{57}$Ti, $^{57}$Sc, and $^{59}$Ti to put the strength of Urca pairs in the $A=57$ and $A=59$ mass chains, and thus for accreting neutron stars with superbursts, on a solid experimental basis. We use total absorption $\gamma$-spectroscopy to avoid the Pandemonium effect and obtain, in connection with measurements of $\beta$-delayed neutron emission, gs-gs branchings. 

\section{Experimental Setup}

The experiment was carried out at the National Superconducting Cyclotron Laboratory at Michigan State University. The experimental technique is the same as described in \cite{ongDecay61VIts2020}. Here, a 140 MeV/u $^{82}$Se primary beam impinged on a 493 mg/cm$^2$ Be target to produce a radioactive ion beam via projectile fragmentation. Fragments were collected by the A1900 fragment separator and selected using the $B\rho-\Delta E-B\rho$ method with a 20~mg/cm$^2$ Kapton wedge placed at the A1900 intermediate dispersive image \cite{morrisseyCommissioningA1900Projectile2003}. The resulting radioactive ion beam with isotopes in the Ca-Cr and $A=53--63$ range was sent to the experimental vault. Beam nuclides were identified event-by-event by measuring energy loss in two PIN Si detectors with 0.503~mm and 1~mm thickness, respectively,  and by measuring time-of-flight between a plastic detector placed near the A1900 intermediate image and the PIN detectors. After passing through an adjustable Al degrader, the mixed beam was implanted in two stations, first for 27 hours in the station surrounded by the NERO neutron long counter \cite{Pereira2010} for the measurement of $\beta$-delayed neutrons, and then for 147 hours into a separate station surrounded by the SuN total absorption $\gamma$-spectrometer \cite{SIMON201316} for the measurement of $\beta$-delayed $\gamma$-rays. 

The NERO station employed the NSCL Beta Counting System \cite{prisciandaro2003beta} and consisted in beam direction of a 1.041~mm Si PIN detector to confirm transmission from the upstream PIN detectors, a 0.997~mm double sided Si strip (DSSD) detector as the main implantation device, a single sided Si-strip (SSSD) detector to veto particles not implanted, and a scintillator to veto light beam particles that mimic $\beta$-particles in the Si detectors. The 4 cm X 4 cm DSSD was segmented in 40x40 strips and signals were split into low and high gain pre-amplifiers to detect the implantation of a heavy ions, as well as subsequent $\beta$ decays, respectively. The DSSD was centered within the NERO long counter detecting neutrons using moderation in a polyethylene matrix, and detection via $^3$He and BF$_4$ neutron counters. The neutron efficiency was determined to be 32(5)\% for neutron energies below 1.5 MeV by scaling the calibration in \cite{Pereira2010} with a $^{252}$Cf source measurement to account for small changes in detector thresholds. 

The SuN station consisted of a 1 mm thick 2 x 2 cm DSSD with 16 x 16 strips for implantation of the beam ions of interest, followed by a Si detector to veto any beam particles passing through the DSSD. The DSSD was centered within the SuN total absorption $\gamma$-spectrometer, consisting of 8 NaI(Tl) scintillators with a high $\gamma$ summing efficiency of 63(2)\% for $^{60}$Co. 

\section{Analysis}
Branchings for $\beta$-delayed neutron emission with NERO were determined as described in \cite{Pereira2010,hosmerHalflivesBranchingsVdelayed2010} from correlating decays detected in the DSSD, both with and without coincident detection of neutrons in NERO,  with a preceding implantation of an ion in space and time. For spatial correlations a 3 x 3 pixel square was used to maximize efficiency. The $\beta$-n coincidence background is determined by using correlations backward in time and is subtracted. 

$\beta$ branchings into $\gamma$-decaying states were determined from SuN data by following largely the approach described in \cite{ongDecay61VIts2020}. GEANT4 simulations were used to create templates of expected total absorption- (TAS, energies of all detector segments summed), single segments- (SS, energies of individual segments merged) and multiplicity (number of segments fired) spectra for the decays of $^{57}$Ti, $^{57}$Sc, and $^{59}$Ti into all energetically possible final states in $^{57}$V, $^{57}$Ti, and $^{59}$V, respectively. Included were the response of SuN for $\gamma$-rays, electrons, and neutrons. For low-lying discrete levels, previous information and SuN data on total absorption, $\gamma$-singles, $\gamma$-$\gamma$ coincidences are used to construct a level scheme with $\gamma$-decay branchings. Higher lying states were treated as a quasi-continuum with templates created for feeding energy bins with a width that corresponds to the detector resolution. The $\gamma$-emission from feeding a bin was determined using the RAINIER software package \cite{kirschRAINIERSimulationTool2018} that uses a level density and a $\gamma$-strength function to follow statistical decays into the discrete level scheme. We used a back-shifted Fermi gas model for the level density, and a standard Lorentzian representation of the E1, M1, and E2 $\gamma$-strength, both with parameters from the TALYS statistical model code \cite{koningTALYSModelingNuclear2023a} for the respective nuclides. Previous experiments have found an upbend in the low energy $\gamma$-strength of nearby $^{57}$Fe \cite{voinovLargeEnhancementRadiative2004}. We therefore include such an upbend in our analysis parametrized as $f_{\rm up}=C \exp(-a E_\gamma)$ with $C=10^{-7}$ MeV$^{-3}$ and $a=1$ MeV$^{-1}$ \cite{gorielyGognyHFB+QRPADipoleStrength2018}. Level feedings are then determined as the obtained weights when fitting the measured single segment, total absorption, and multiplicity spectra as a linear combination of all templates. Prior to fitting, the experimental spectra were corrected by subtracting background and contributions from daughter decays, which were not included in the simulation. Systematic errors were determined by performing the fit 10$^6$ times using a Multi-Objective Evolutionary Algorithm based on Decomposition with a Differential Evolver (MOEA/D-DE) \cite{zhangMOEAMultiobjectiveEvolutionary2007a} (for parameters see \cite{supplemental_material}). This approach explores the tradeoffs that can be made when fitting multiple objectives (the different spectra) with imperfect templates. For each fit, a summed $\chi ^2$ from all spectra was determined. The distribution of solutions that produced a $\chi ^2$ within a 1-$\sigma$ range resulted in corresponding distributions for each level feeding, from which a 1$\sigma$ error was extracted. Statistical errors were determined by generating and then analyzing bin-by-bin random variations of the measured spectra using a Monte Carlo approach.

\section{Results}
\subsection{$P_{\rm n}$ values}
\begin{figure}[tbp]
\includegraphics[width=0.95\linewidth]{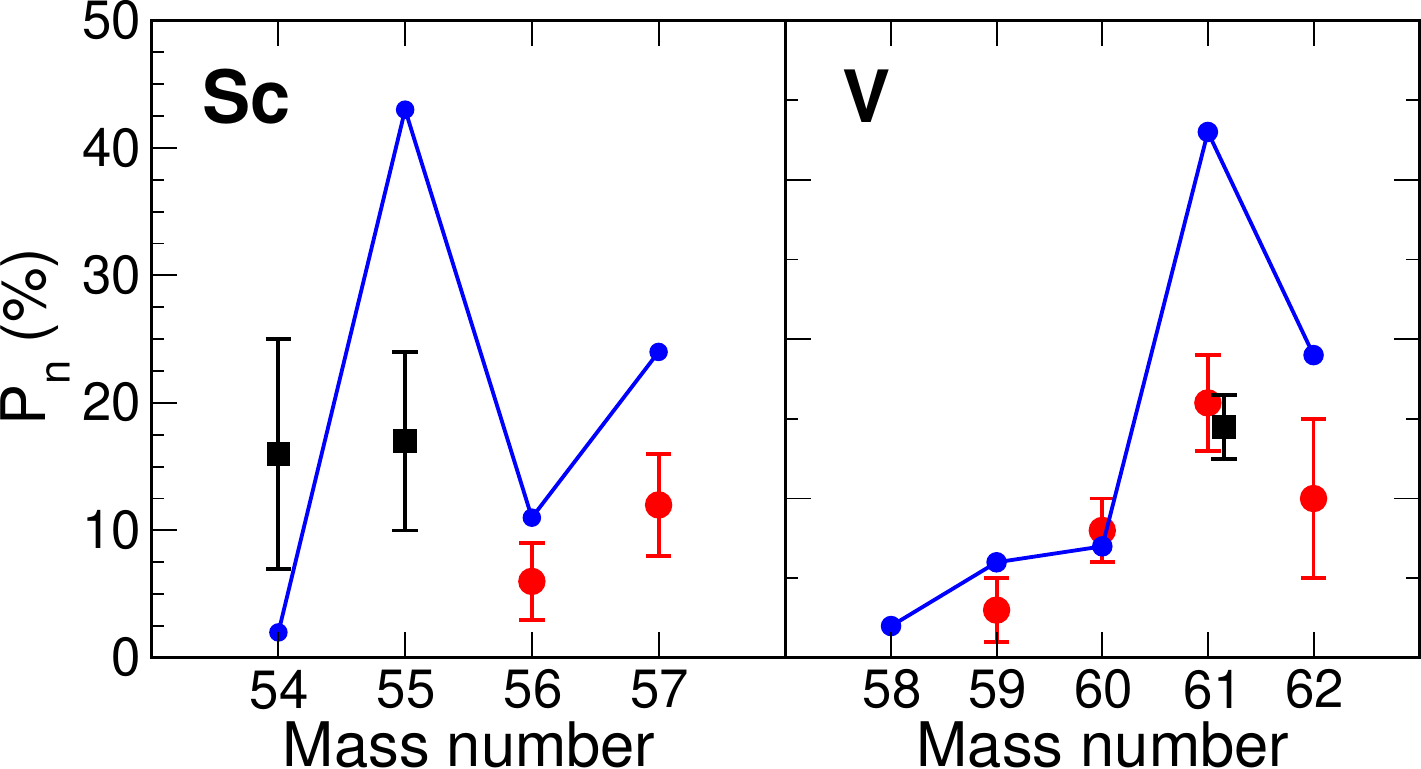}
\caption{$P_{\rm n}$ values as functions of mass number for Sc isotopes (left) and V isotopes (right). Shown are the results from this work (red), results from previous work (black) \cite{crawfordDecayIsomericProperties2010b,ongDecay61VIts2020}, and theoretical predictions from QRPA-fY (blue). 
\label{fig:pn}}
\end{figure}

Tab.~\ref{tab:pn} lists the $P_{\rm n}$ values obtained from the measurements with NERO for the implanted ions. These values are then used in the further template analysis. The only previously reported $P_{\rm n}$ value for $^{61}$V of 14.5(2.0)\% is in good agreement with our value of 16(4)\%. In Fig.~\ref{fig:pn} we compare the new results with predictions from the QRPA-fY theory \cite{mollerNewDevelopmentsCalculation1990,mollerNuclearPropertiesAstrophysical1997,Moller2019} combined with a statistical model \cite{mumpowerNeutrongCompetitionDelayed2016a}. Overall there is reasonable agreement for a global model, though theory appears to overpredict the $P_{\rm n}$ values somewhat.

\begin{table}
\caption{\label{tab:pn} $P_{\rm n}$ values and half-lives obtained in this work compared to literature values.}
\begin{ruledtabular}
\begin{tabular}{cccccc}
           & \multicolumn{2}{c}{$P_{\rm n}$ (\%)} & \multicolumn{2}{c}{$T_{12}$ (ms)} \\
Isotope    & This work & Literature & This work & Literature \\  
\hline
$^{56}$Sc\footnote{May be mixture of ground state and high-spin isomer \cite{crawfordDecayIsomericProperties2010b}}  & 6(3)  &                                    &       & \\
$^{57}$Sc  & 12(4) &                                    & 20(1) & 13(4) \cite{gaudefroyBetadecayStudiesNeutronrich2005a} \\
           &         &                                    &        & 22(2) \cite{crawfordDecayIsomericProperties2010b} \\
$^{57}$Ti  & $<$5    &                                    & 94(1) & 98(5) \cite{liddickVdecayOddA57Ti2005}\\
$^{58}$Ti  & $<$3    &                                    & 52(1) & 59(9) \cite{gaudefroyBetadecayStudiesNeutronrich2005a} \\
$^{59}$Ti  & 4(2)  &                                    & 26(1) & 30(3) \cite{gaudefroyBetadecayStudiesNeutronrich2005a}\\
$^{59}$V   & 3(2)  &                                    & 91(2) & 97(2) \cite{liddickVdecayOddA57Ti2005}\\
$^{60}$V   & 8(2)  &                                    & 77(1) & 68(4) \cite{sorlinBetaDecayStudies2003}\\
$^{61}$V   & 16(3) & 14.5(2.0) \cite{ongDecay61VIts2020} & 46(1) & 47(1) \cite{sorlinBetaDecayStudies2003}\\
           &       &                                    &       & 48(1) \cite{ongDecay61VIts2020}\\
$^{62}$V   & 10(5) &                                    & 32(2) & 33(2) \cite{sorlinBetaDecayStudies2003}\\
\end{tabular}
\end{ruledtabular}
\end{table}

\subsection{$^{57}$Ti Decay}
\begin{figure}[tbp]
\includegraphics[width=0.95\linewidth]{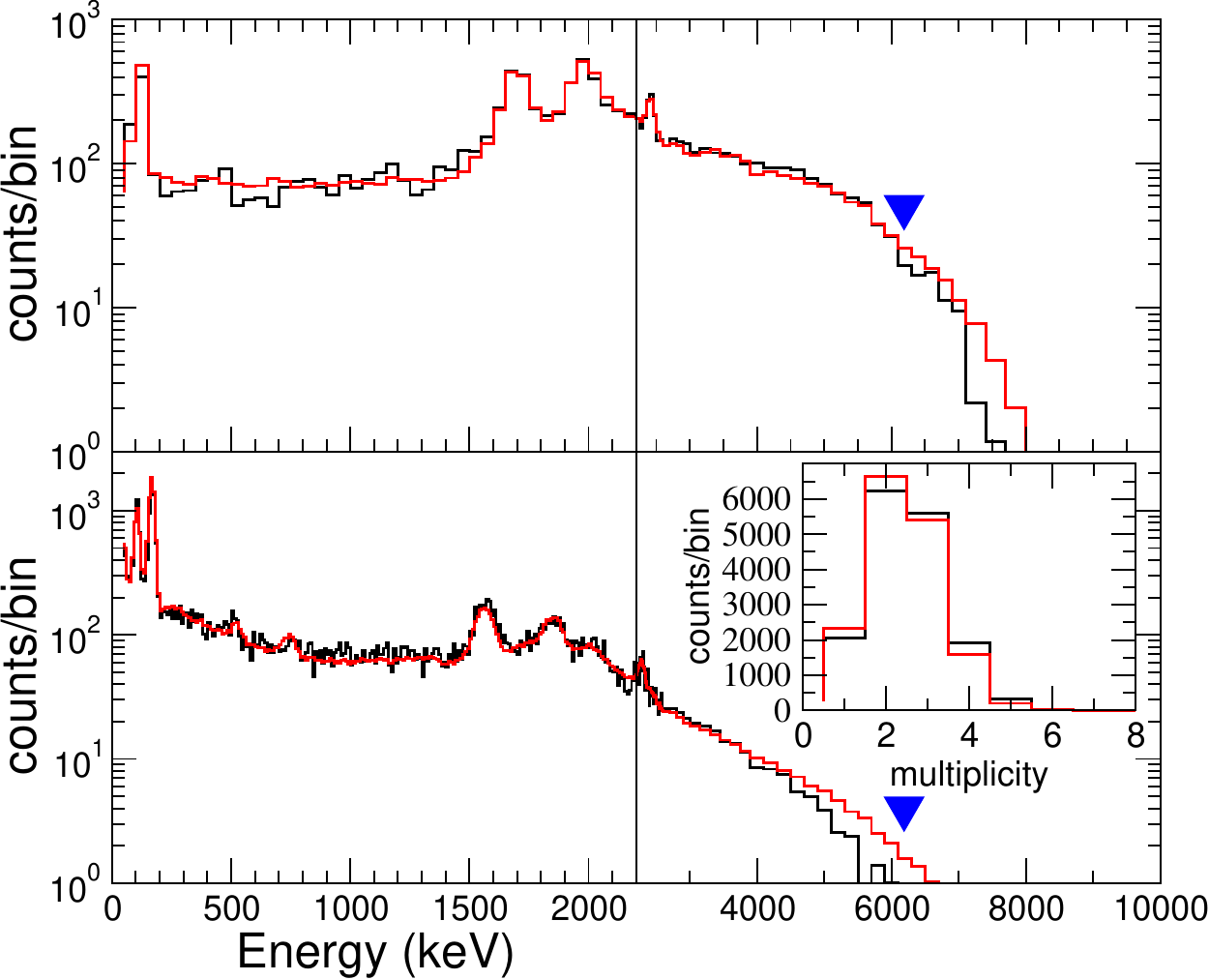}
\caption{TAS (top), SS (bottom), and multiplicity (bottom inset) measured spectra (black) together with the corresponding template fits (red) for the decay of $^{57}$Ti. The black vertical line indicates a change in the energy axis scale. 
Blue triangles mark the neutron separation energy of the $^{57}$V daughter. 
\label{FigSpectraTi57}}
\end{figure}

\begin{figure}[tbp]
\includegraphics[width=0.85\linewidth]{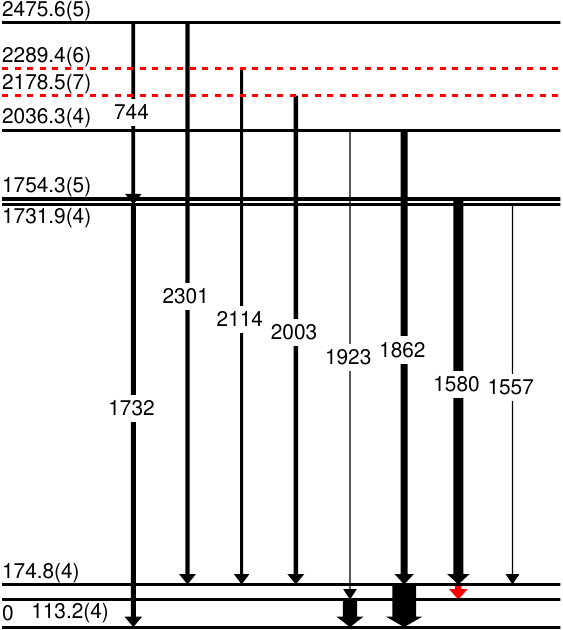}
\caption{Discrete levels in $^{57}$V populated by the decay of $^{57}$Ti and associated $\gamma$-transitions used in this work. The newly identified transition is drawn in red. Dashed red lines indicate new tentative levels that have been inserted to improve the fits.  
\label{fig:57ti_level}}
\end{figure}

Fig.~\ref{FigSpectraTi57} shows the total absorption, singles, and multiplicity spectra together with the resulting best template fit for the decay of $^{57}$Ti. For the discrete level scheme, we use all levels and $\gamma$-branches from \cite{liddickVdecayOddA57Ti2005}. Signatures for all previously reported levels are observed in the TAS spectrum, though the 1731.9 and 1754.3 keV levels are not resolved. We do find evidence for 2 additional levels located between the previously reported 2036.3 and 2475.6 keV levels. When gating on a count rate excess in the 2100--2300 keV region in the total absorption spectrum we observe evidence for the two previously observed but unplaced 2003.7 and 2114.6 keV $\gamma$-rays \cite{liddickVdecayOddA57Ti2005} in coincidence with 175 keV $\gamma$-rays. We therefore tentatively place these two unplaced $\gamma$-rays  on top of the 174.8~keV state, resulting in new levels at 2178 and 2289 keV (Fig.~\ref{fig:57ti_level}). While not resolved, these two levels lead to a significantly improved fit of the total absorption spectrum in this energy region. We also observe the 62~keV $\gamma$-ray from the decay of the 174.8~keV to the 113.2~keV level that was inferred but below the detection energy threshold in \cite{liddickVdecayOddA57Ti2005}. 

The resulting level scheme is shown in Fig.~\ref{fig:57ti_level} and $\beta$-feeding intensities are listed in Tab.~\ref{TabFeeding57Ti}. Asymmetric uncertainties for the intensities have been symmetrized following \cite{nubase1997}. Compared to the previous results from high-resolution $\gamma$-ray  spectroscopy \cite{liddickVdecayOddA57Ti2005} we find significantly larger feeding of excited states. As a consequence, the inferred ground-state feeding is dramatically reduced from 54(3)\% to 4(2)\% (Tab.~\ref{TabFeeding57Ti}). Log$ft$ values were calculated using a decay Q-value of 9.98(21)~MeV obtained using the AME20 \cite{wangAME2020Atomic2021} atomic mass of $^{57}$Ti  (-34.402(206)~MeV) as well as the weighted average of two more recent precision mass measurements of $^{57}$V (-44.377(11)~MeV) obtained with the MR-TOF \cite{iimuraStudy$N32$$N34$2023} and Penning Trap \cite{porterInvestigatingNuclearStructure2022} techniques, respectively. We use our half-life of 94(1)~ms that is significantly more precise than previous measurements.  The resulting log$ft$ for the ground-state transition is 5.8$^{+0.3}_{-0.2}$, with the error dominated by the error in the relatively small branching ratio. 

\begin{table}
\caption{\label{TabFeeding57Ti} Relative $\beta$-feeding intensities for the discrete states in $^{57}$V from the decay of $^{57}$Ti}
\begin{ruledtabular}
\begin{tabular}{ccc}
Energy (MeV) & Feeding (\%) & log$ft$ \footnote{For asymmetric error bars, the two values in brackets indicate uncertainty on last digits up and down, respectively} \\
\hline
0 & 4(2) & 5.8(3,2)\\
113.2 & 0 & \\
174.8 & 9(3) & 5.3(2,1)\\
1732\footnote{States not resolved in this work} & \multirow{2}{*}{25(3)} & \multirow{2}{*}{4.50(8)} \\
1754$^a$ &  \\
2036 & 26(4) & 4.40(9)\\
2178 & 4(2) & 5.2(3,2)\\
2289 & 3(1) & 5.3(2,1)\\
2475 & 12(3) & 4.6(1) \\
\end{tabular}
\end{ruledtabular}
\end{table}

\subsection{$^{57}$Sc decay}
\begin{figure}[tbp]
\includegraphics[width=0.95\linewidth]{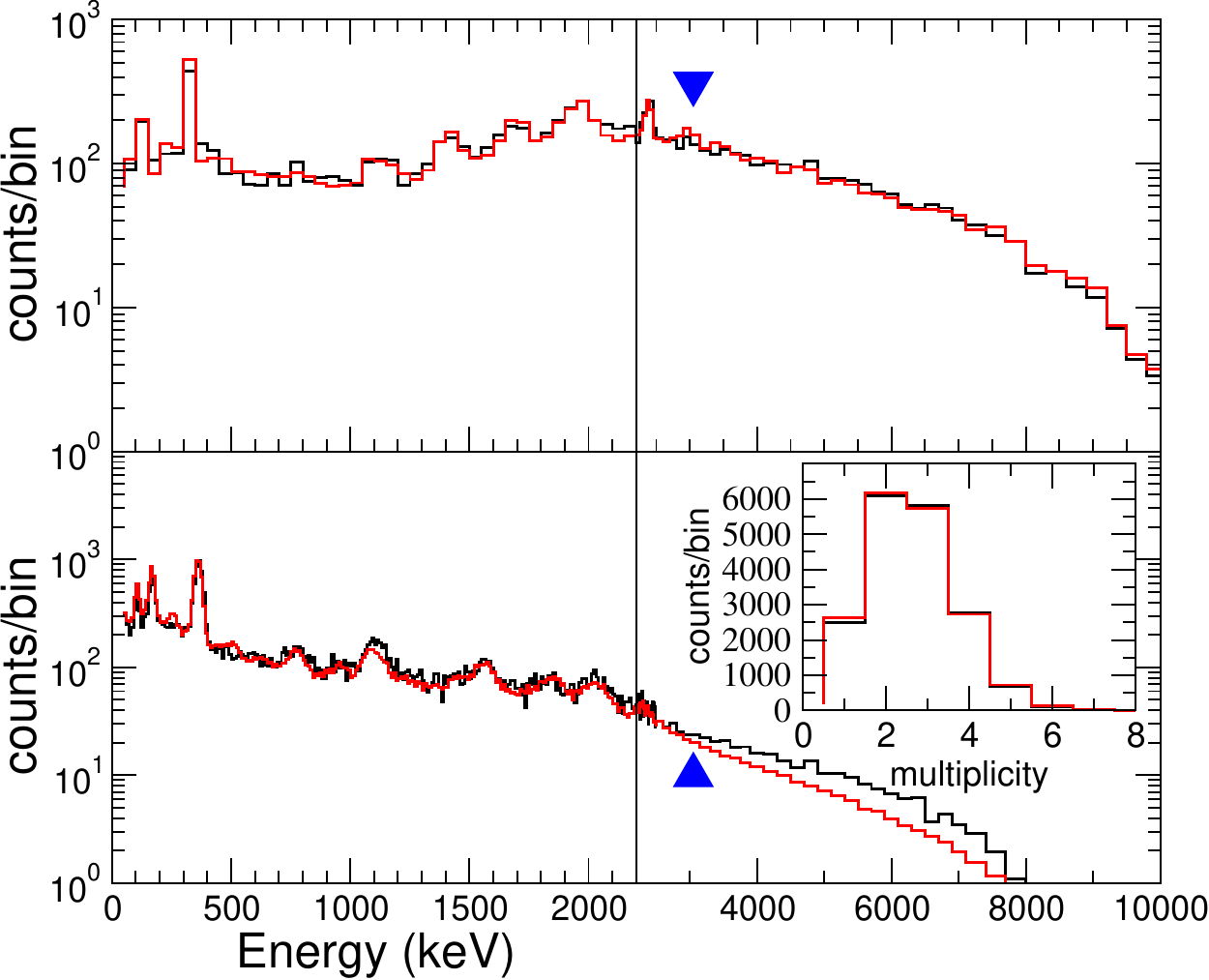}
\caption{TAS (top), SS (bottom), and multiplicity (bottom inset) measured spectra (black) together with the corresponding template fits (red) for the decay of $^{57}$Sc. The vertical black line marks a change in energy axis scale. Blue triangles mark the neutron separation energy of the $^{57}$Ti daughter. 
\label{FigSpectraSc57}}
\end{figure}

\begin{figure}[tbp]
\includegraphics[width=0.85\linewidth]{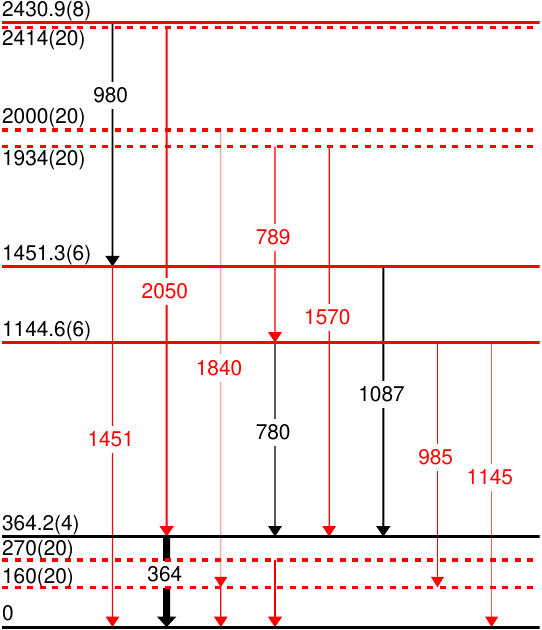}
\caption{Discrete levels in $^{57}$Ti populated by the decay of $^{57}$Sc and associated $\gamma$-transitions used in this work. Newly identified levels and transitions are drawn in red. Dashed lines indicate tentative levels that have been inserted to improve the fits.  
\label{fig:57sc_level}}
\end{figure}

Fig.~\ref{FigSpectraSc57} shows the total absorption, singles, and multiplicity spectra together with the resulting best template fits for the decay of $^{57}$Sc. \cite{crawfordDecayIsomericProperties2010b} reported evidence for a 364~keV first excited state, and 4 unplaced $\gamma$-rays from $^{57}$Ti from high-resolution $\gamma$-spectroscopy. Using our total absorption spectrum and $\gamma$-$\gamma$ coincidences we can place 3 of these into a level scheme (Fig.~\ref{fig:57sc_level}). We do not see evidence for the 1127 keV $\gamma$-ray reported in \cite{crawfordDecayIsomericProperties2010b} in the $^{57}$Ti level scheme, supporting their assignment of this transition to the $\beta$-delayed neutron daughter. Indeed their reported 12\% intensity for this $\gamma$-ray is consistent with our measured $P_{\rm n}=12(4)$ \%. We do observe 2 additional $\gamma$-rays that that we can assign based on these levels and for which we can thus infer accurate energies indirectly using the data from \cite{crawfordDecayIsomericProperties2010b}. We observe 6 additional $\gamma$-rays with energies of 160, 820, 984, 1570, 1840, and 2050~keV that based on $\gamma$-$\gamma$- coincidences we place tentatively into 5 additional levels. Owing to the limited resolution of SuN, the inferred energies of these level have large uncertainties of about 20~keV. Although the level pairs around 2.0 and 2.4 MeV could also be a single state, we do treat them separately for the purpose of this analysis. A count excess around 270~keV requires an additional level to improve the fit to the data around that energy. We include such a level in the analysis, though uncertainties in this region are large owing to relatively large daughter subtraction, resulting in a large uncertainty of the inferred feeding into this region. The existence of this level is therefore very uncertain. 

\begin{table}
\caption{\label{TabFeeding57Sc} Relative $\beta$-feeding intensities for the discrete states in $^{57}$Ti from the decay of $^{57}$Sc}
\begin{ruledtabular}
\begin{tabular}{ccc}
Energy (keV) & Feeding (\%) & log$ft$ \\
\hline
0 & $<$0.9 & $>$6.0\\
160 & $<$0.6 & $>$6.1 \\
270 & 3(1) & 5.6(2)\\
364.2 & 23(5) & 4.7(1) \\
1144.6 & $<$2 & $>$5.1\\
1451.3 & 7(1) & 5.1(1)\\
$\approx$2000 & 6(3) \\
$\approx$2400 & 13(3) \\
\end{tabular}
\end{ruledtabular}
\end{table}

The $\beta$-decay templates for $^{57}$Sc included $\beta$-delayed neutron emission to the ground state and the known first three excited states in $^{56}$Ti \cite{liddickDevelopmentShellClosures2004}. During the fitting, the measured total $P_{\rm n}=12$ \% from this work was held fixed, while the branchings into the different final states were free parameters. We only saw evidence for feeding of the first excited state in $^{56}$Ti at 1.128~MeV. 
The resulting $\beta$-feeding intensities in $^{57}$Ti are listed in Tab.~\ref{TabFeeding57Sc}. The spectra can be well fitted without a ground-state transition, for which an upper limit of 0.9\% is obtained. While the 160~keV state could also play a role in Urca cooling at high temperatures, its feeding is similarly small. Logft values were calculated with a Q-value of -13.02(27)~MeV using AME2020 \cite{wangAME2020Atomic2021} masses based on time-of-flight mass measurements \cite{michimasaMappingNewDeformation2020,meiselNuclearMassMeasurements2020}. We use the half-life of 20(1)~ms from this work, which agrees with the previously obtained half-life from \cite{crawfordDecayIsomericProperties2010b} but is more precise. We obtain a lower limit of the gs-gs log$ft$ of 6.0. 

\subsection{$^{59}$Ti decay}
\begin{figure}[tbp]
\includegraphics[width=0.5\linewidth]{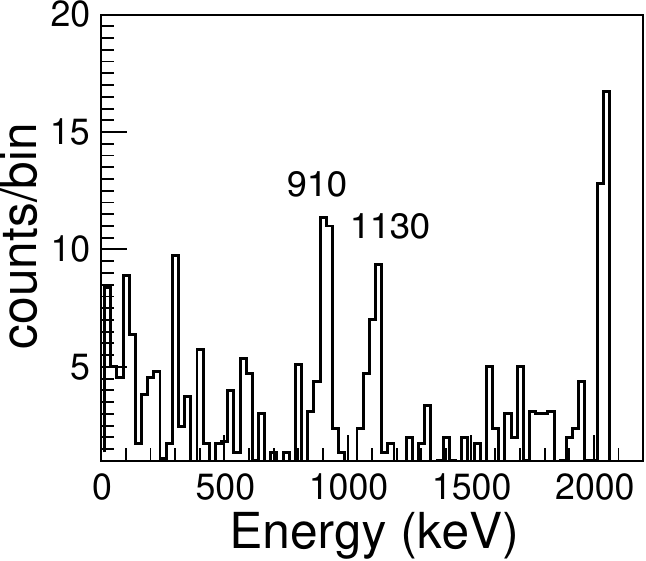}
\caption{$\gamma$-$\gamma$ single segment (SS) coincidence spectrum, gated on a TAS energy of 2025--2075~keV.  
\label{fig:ti59_gg}}
\end{figure}

\begin{figure}[tbp]
\includegraphics[width=0.95\linewidth]{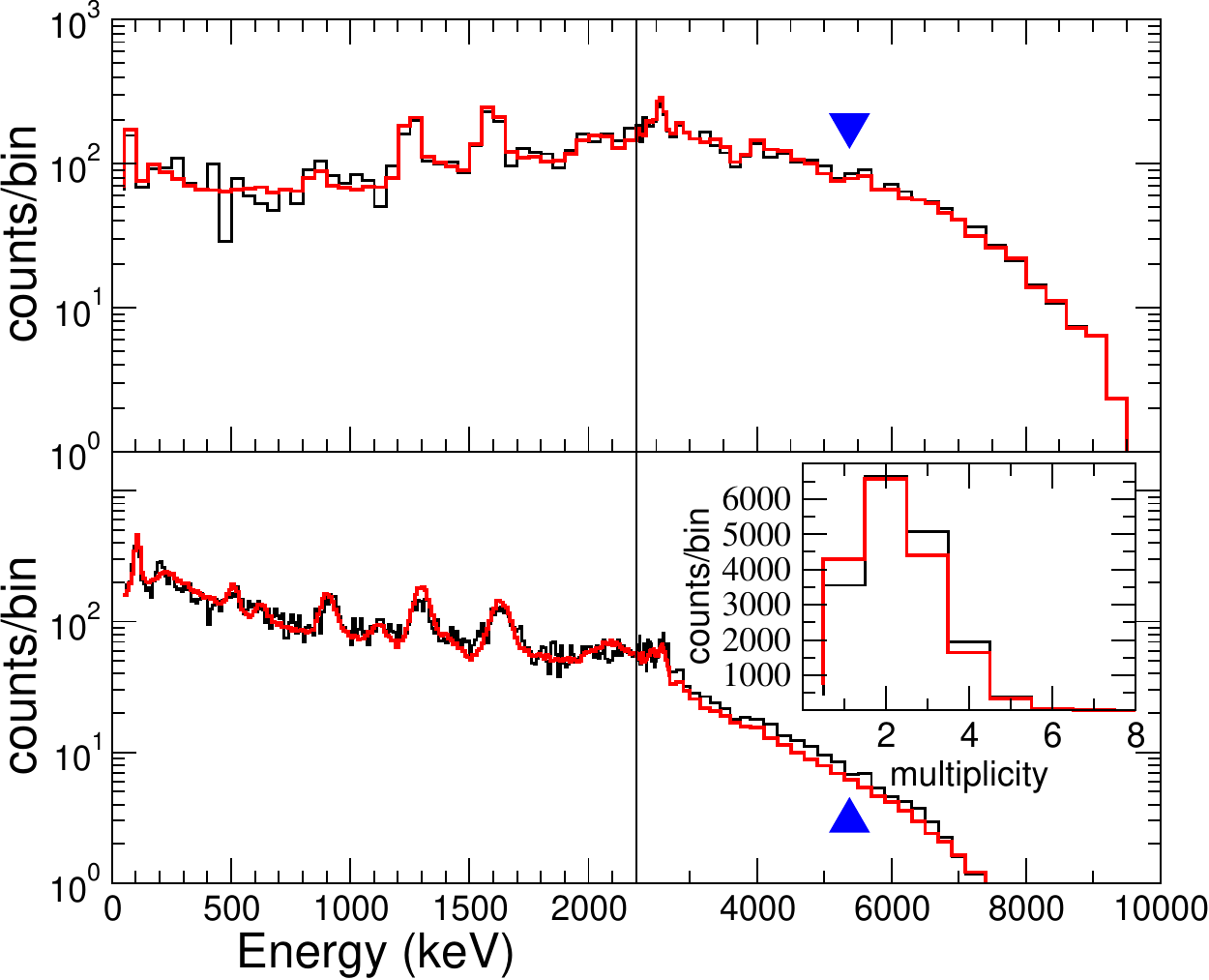}
\caption{TAS (top), SS (bottom), and multiplicity (bottom inset) measured spectra (black) together with the corresponding template fits (red) for the decay of $^{59}$Ti. The vertical black line marks a change in energy axis scale. Blue triangles mark the neutron separation energy of the $^{59}$V daughter. 
\label{FigSpectraTi59}}
\end{figure}

\begin{figure}[tbp]
\includegraphics[width=1.0\linewidth]{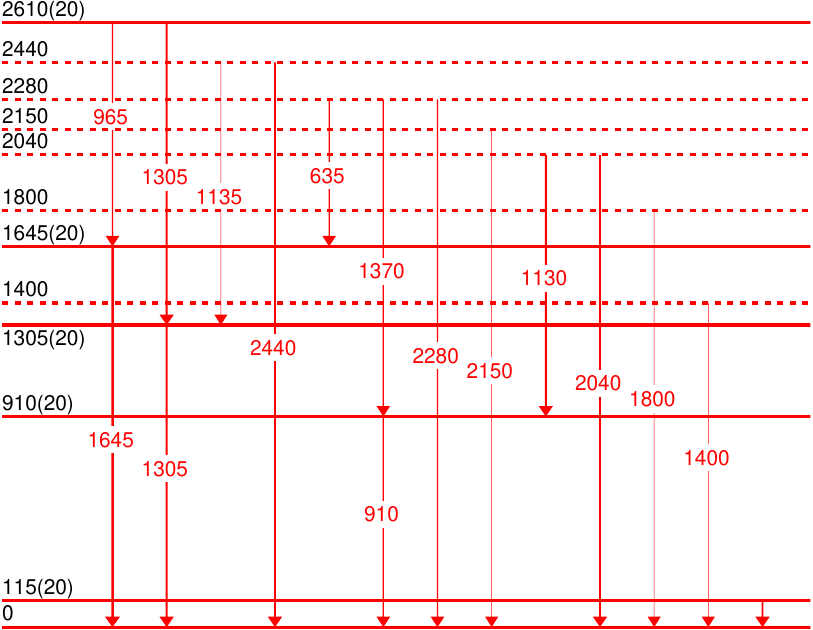}
\caption{Discrete levels in $^{59}$V populated by the decay of $^{59}$Ti and associated $\gamma$-transitions used in this work. All levels and transitions are newly identified and thus drawn in red. Dashed lines indicate tentative levels that have been inserted to improve the fits. 
\label{fig:59ti_level}}
\end{figure}

Fig.~\ref{FigSpectraTi59} shows the total absorption, singles, and multiplicity spectra together with the resulting best template fits for the decay of $^{59}$Ti. Previous information on $\gamma$-spectroscopy of $^{59}$V is limited to the detection of three $\gamma$-rays with energies of 910, 1051, and 1135~KeV using the $^{62}$V(p,p2n)$^{59}$V reaction \cite{ elekesSouthwesternBoundary402022}, which does not necessarily populate the same levels as the $^{59}$Ti $\beta$ decay. A proposed level scheme is based on systematics and
shell-model predictions only. We therefore construct a level scheme solely from our data (Fig.~\ref{fig:59ti_level}).  The TAS spectrum shows clear evidence of levels at 115, 1305, 1645, and 2610~KeV, and a smaller peak indicating a level at 910~KeV. Three additional levels were required in the 2000--2500 keV range to explain the strong feeding in that energy region and are marked as tentative. Level energies, and decay schemes were constructed from $\gamma$-$\gamma$ coincidence analysis. We do observe two of the three previously reported $\gamma$-rays at around 910 and 1130 keV. However, we find a clear coincidence of both with a TAS gate around 2040 keV and no additional $\gamma$-rays (Fig.~\ref{fig:ti59_gg}). We therefore place these $\gamma$-rays into a cascade from a 2040~keV excited state to the ground state. This differs from the placement in \cite{elekesSouthwesternBoundary402022} though the $\gamma$-rays observed here may not necessarily be the ones observed in $^{62}$V(p,p2n)$^{59}$V. The 115~keV state may be the 5/2$^-$ 152~keV state predicted by the shell model but not observed in \cite{ elekesSouthwesternBoundary402022}. We added 3 states at 1400~keV, 1800~keV, and 2150~keV to correct small deficiencies in the TAS spectrum just above the large peaks at 1305~keV, 1645~keV, and 2040 keV. This also led to an improvement of the SS spectrum and resulted in a small change, within error bars, of the inferred ground-state feeding. The resulting $\beta$-feeding intensities are listed in Tab.~\ref{TabFeeding59Ti}. The ground-state feeding is small with 7.7(10)\%, but not negligible. Log$ft$ values were calculated with a Q-value of $Q=12.40(22)$~MeV, obtained using the AME2020 \cite{wangAME2020Atomic2021} mass for $^{59}$V and for $^{59}$Ti the weighted average -25.21(18)~MeV of the two independent TOF mass measurements at RIKEN RIBF \cite{michimasaMappingNewDeformation2020} and MSU \cite{meiselNuclearMassMeasurements2020} that are in good agreement. We use a half-life of 26(1)~ms obtained in this work. The resulting gs-gs log$ft$ value is 5.34$^{+0.08}_{-0.24}$.

\begin{table}
\caption{\label{TabFeeding59Ti} Relative $\beta$-feeding intensities for the discrete states in $^{59}$V from the decay of $^{59}$Ti}
\begin{ruledtabular}
\begin{tabular}{ccc}
Energy (keV) & Feeding (\%) \footnote{For asymmetric error bars, the two values in brackets indicate uncertainty on last digits up and down, respectively} & log$ft$ \footnotemark[\value{footnote}]\\
\hline
0 & 6.9 (40,6) & 5.34(8,24)\\
115 & 2.8(11,4) & 5.7(1,2)\\
1305 & 7.3(13,8) & 5.1(1)\\
1645 & 9.4(16,9) & 4.9(1) \\
2040 & 2.6(5,9) & 5.4(2,1)\\
2280 & 4.9(10,16) & 5.1(2,1)\\
2440 & 6.1(2,3) & 4.95(7) \\
2610 & 13.1(20,11) & 4.6(1) \\
\end{tabular}
\end{ruledtabular}
\end{table}

\section{Discussion}
\begin{figure}[tbp]
\includegraphics[width=0.95\linewidth]{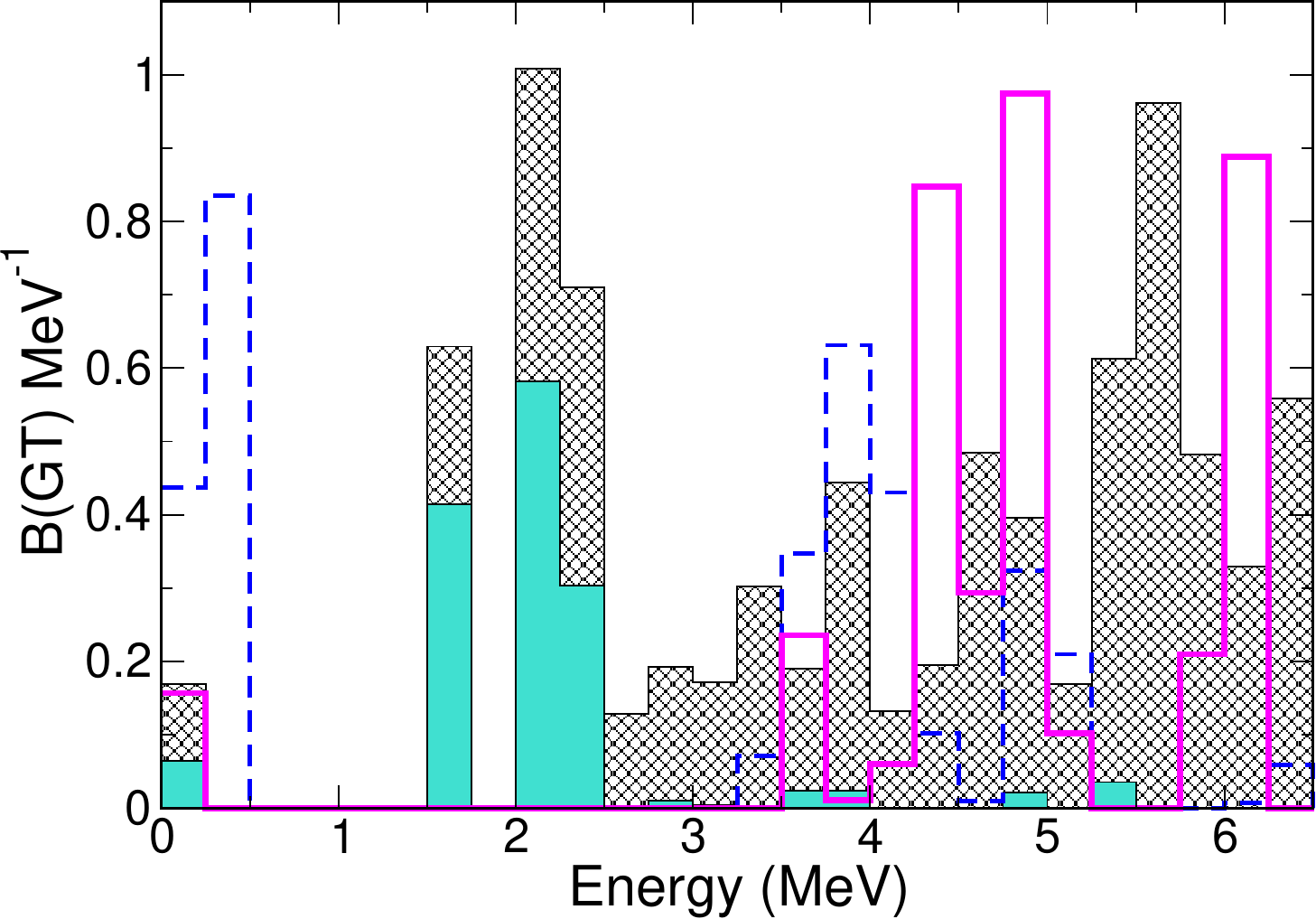}
\caption{B(GT) $^{57}$Ti $\beta$-decay strength as function of excitation energy in $^{57}$V. Shown are experimental results (green filled histogram with grey patterned uncertainty range), the QRPA-fY theoretical default predictions (blue, dashed) with $\epsilon_2=0.12$, and a QRPA-fY calculation with the initial unpaired neutron placed in a 1/2$^-$ Nilsson orbital (magenta, solid).  
\label{fig:strength_ti57}}
\end{figure}

\begin{figure}[tbp]
\includegraphics[width=0.95\linewidth]{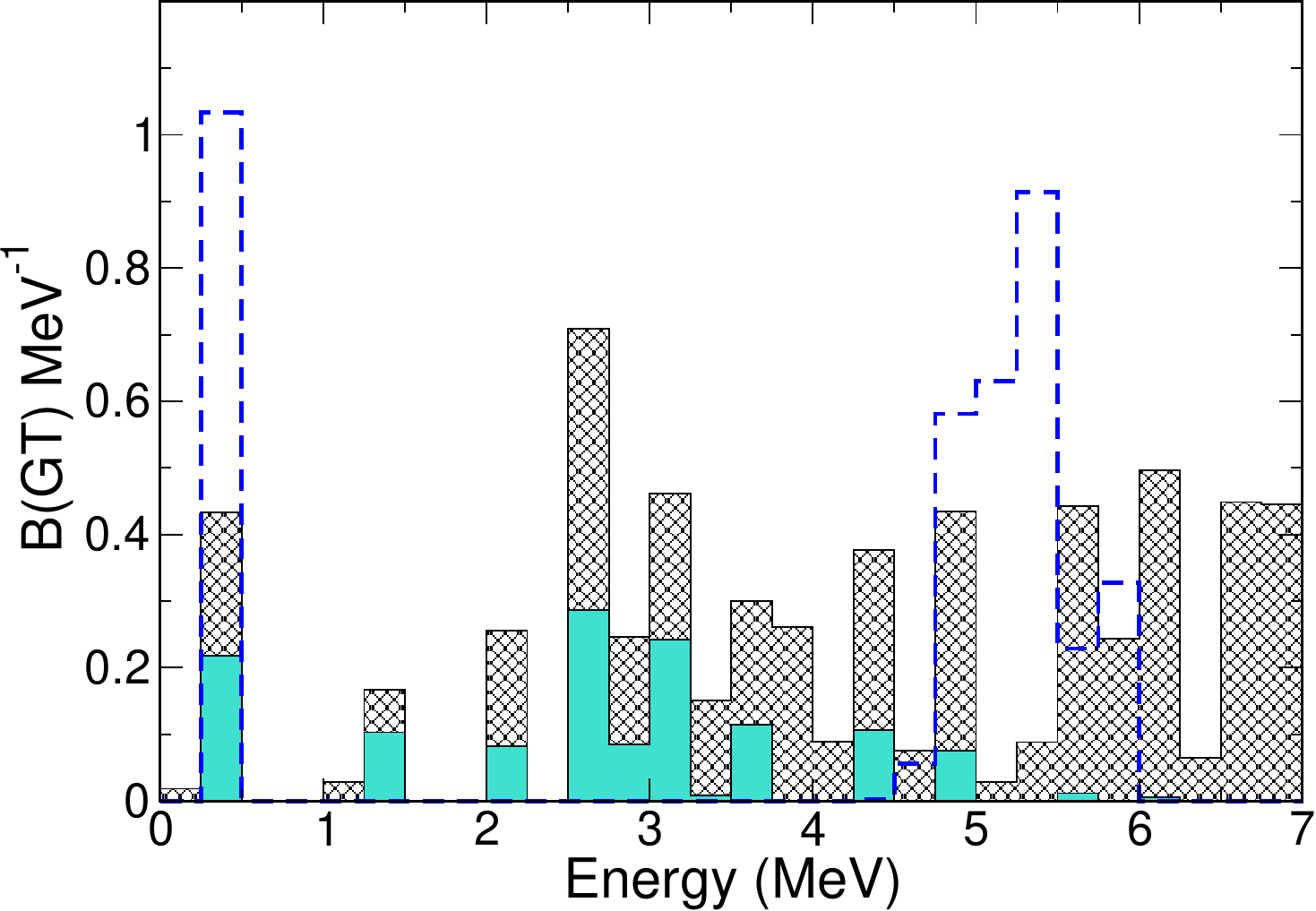}
\caption{$^{57}$Sc $\beta$-decay strength as function of excitation energy in $^{57}$Ti. Shown are experimental results (green filled histogram with grey patterned uncertainty range), and the QRPA-fY theoretical default predictions with $\epsilon_2=$-0.1 (blue, dashed). 
\label{fig:strength_sc57}}
\end{figure}

\begin{figure}[tbp]
\includegraphics[width=0.95\linewidth]{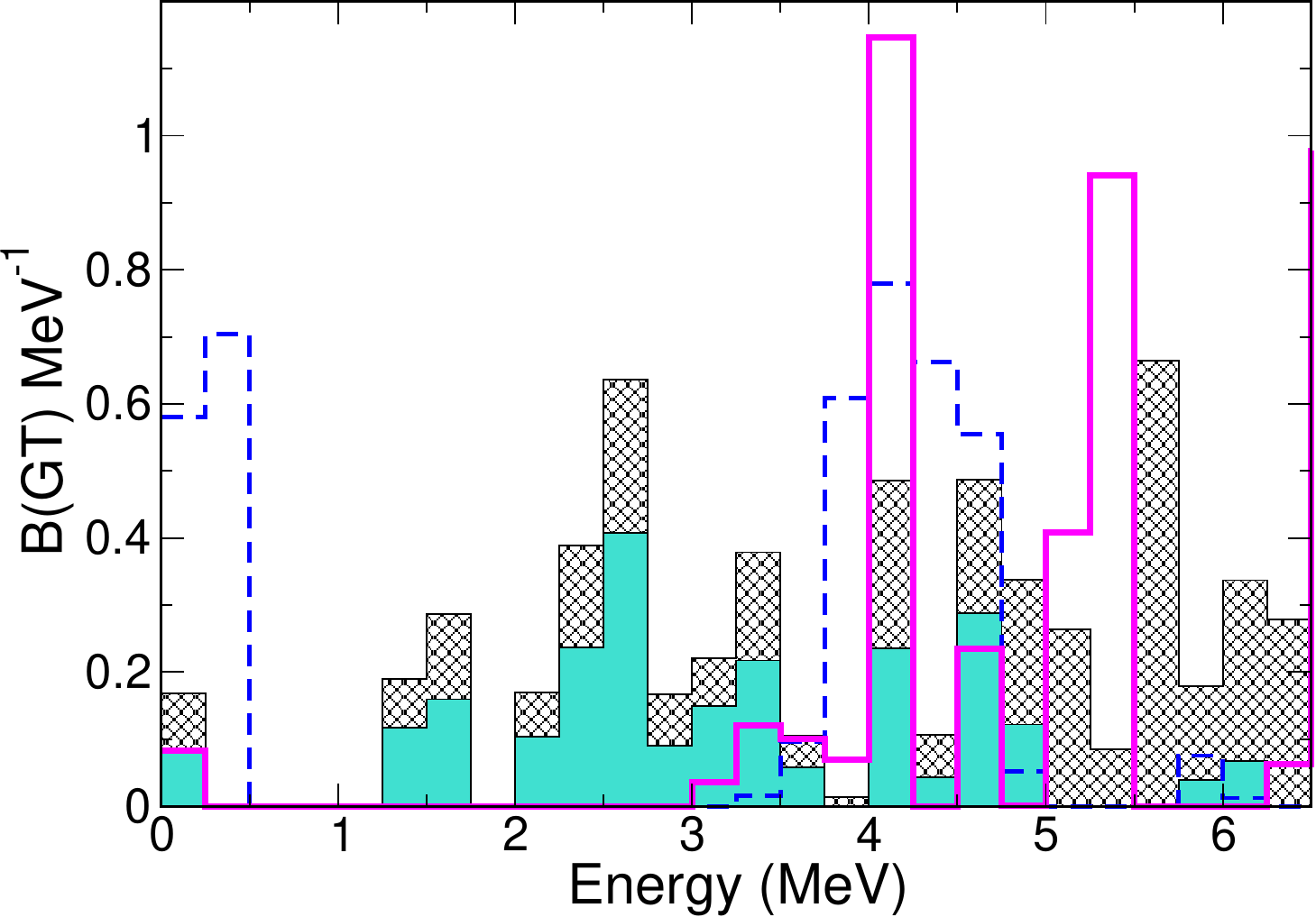}
\caption{$^{59}$Ti $\beta$-decay strength as function of excitation energy in $^{59}$V. Shown are experimental results (green filled histogram with grey-patterned uncertainty range), the QRPA-fY theoretical default predictions with $\epsilon_2=$-0.1 (blue, dashed), and a QRPA-fY calculation with $\epsilon_2=$-0.15, which results in the initial unpaired neutron being placed in a 1/2$^-$ Nilsson orbital (magenta, solid).  
\label{fig:strength_ti59}}
\end{figure}

Figs.~\ref{fig:strength_ti57},\ref{fig:strength_sc57}, and \ref{fig:strength_ti59} show B(GT) strength distributions obtained from the feeding intensities determined in this work, compared with the global QRPA-fY model calculations \cite{mollerNewDevelopmentsCalculation1990,mollerNuclearPropertiesAstrophysical1997,Moller2019} typically used in neutron star crust model calculations. In qualitative agreement with the experimental data, QRPA-fY predicts in all three cases significant strength below 0.5~MeV and in the 4--6~MeV range. In the 4--6~MeV range there is some consistency between experiment and theory though experimental error bars are very large. For the $^{57}$Sc case the neutron separation energy of $^{57}$Ti is 3.0~MeV and thus the 4-6~MeV strength appears not in the $\gamma$-data but in the $P_{\rm n}$ value. The QRPA prediction of $P_{\rm n}$=20\% is in reasonable agreement with our measured 12(4)\%. For the $<$0.5~MeV strength of main interest here, experiment and theory agree that for the $^{57}$Ti and $^{59}$Ti decays this strength is in the lowest energy bin consistent with the ground state, while for the $^{57}$Ti decay it is located in the 0.25--0.5~MeV bin. However, in all three cases we find experimentally a much smaller $<$0.5~MeV strength than predicted. Instead experiment shows significant strength in the 1--3~MeV range that is not predicted by theory. 


Nuclear deformation significantly affects $\beta$-decay properties. The QRPA-fY model uses deformations predicted by the FRDM12 mass model \cite{MOLLER20161}, which predicts very small deformations $-0.1 < \epsilon_2  < 0.1$ for $^{57}$Sc, $^{57}$Ti, and $^{59}$Ti and spherical shapes for more neutron-rich Ti isotopes out to $N=40$ $^{62}$Ti. However, there is strong experimental evidence for the emergence of significant deformation around $N=40$ between Ca and Ni \cite{cortesShellEvolution402020} giving rise to a new so called island of inversion in this region \cite{lilianacortesIslandInversion$N40$2024, nowackiNeutronrichEdgeNuclear2021, elekesSouthwesternBoundary402022}. For the Ti isotopic chain, systematics of the energy of the first 2$^+$ state ($E(2^+)$) indicates the onset of moderate but significant deformation at $^{60}$Ti that increases to $\epsilon_2 \approx 0.3$ at $^{62}$Ti \cite{cortesShellEvolution402020}. 

While deformation is unlikely to affect the calculations of $^{57}$Sc and $^{57}$Ti decays, it is possible that $^{59}$Ti is already part of the island of inversion and significantly deformed. To explore this possibility, we carried out QRPA-fY calculations for the decay of $^{59}$Ti with increasing quadrupole deformation. As deformation increases we find, as expected, an increasing role played by the 1/2$^+$[440] Nilsson orbital related to the spherical g9/2 shell, which dramatically lowers in energy and moves closer to the Fermi surface. As a consequence, the predicted gs-gs strength systematically decreases with increasing deformation and vanishes completely at $\epsilon_2=0.2$. Thus, moderate deformation of $^{59}$Ti may indeed explain the overestimation of the experimental gs-gs strength by theory. 

Spherical shell-model calculations can also shed light on the interpretation of our results. GXPF1 fp-shell model calculations for the decay of $^{57}$Ti indicate a very strong transition from a 5/2$^-$ gs to a 7/2$^-$ gs \cite{liddickVdecayOddA57Ti2005}, in line with the QRPA predictions, but in disagreement with our new experimental results. Recently the LEPS shell model built on a $^{48}$Ca core, and including the full fp-shell, as well as g9/2 and d5/2 for neutrons, has been successful in reproducing experimental data in this region on the nuclear chart, for example knockout cross sections in $^{60}$Ti \cite{gadeNuclearStructure$N40$2014}. A recent calculation with this model led to proposed revised spin assignments for $^{57}$Ti and $^{59}$Ti (and $^{61}$Ti), with 1/2$^-$ gs and 5/2$^-$ first excited states \cite{wimmerFirstSpectroscopy61Ti2019}. The authors cite the strong feeding of the second excited state in $^{57}$Ti qualitatively observed in the $\beta$ decay of 7/2$^-$ $^{57}$Sc with high-resolution $\gamma$-spectroscopy \cite{crawfordDecayIsomericProperties2010b} as evidence. Our measurement of the $^{57}$Sc decay provides quantitative feeding data for the first time and confirm this result, with the first excited 364~KeV state in $^{57}$Ti having the strongest feeding with 21\%, while the feeding of the ground state is compatible with zero (Tab.~\ref{TabFeeding57Sc}). We explored the possible impact of a 1/2$^-$ parent state in the QRPA-fY calculations of the decays of $^{57}$Ti and $^{59}$Ti by placing the initial unpaired neutron in the 1/2$^-$[301] Nilsson orbital emerging from the p1/2 spherical shell level. While for $^{57}$Ti we had to place the neutron artificially into that orbital, for $^{59}$Ti QRPA-fY predicts this level to be the ground state for a quadrupole deformation of $\epsilon_2=0.15$. We therefore simply chose that deformation. The resulting calculations show significantly better agreement with experimental data for strength below 0.25~MeV excitation energy (Figs.~\ref{fig:strength_ti57} and \ref{fig:strength_ti59}). 

\section{Astrophysical Implications}
The inferred transition strengths to low-lying states in the $\beta$ decays of $^{57}$Sc, $^{57}$Ti, and $^{59}$Ti directly inform the rate of Urca cooling in neutron star crusts via the $^{57}$Ti$ \leftrightarrow ^{57}$Sc, $^{57}$V$ \leftrightarrow ^{57}$Ti, and $^{59}$V$ \leftrightarrow ^{59}$Ti electron capture/$\beta$-decay Urca pairs. To explore the astrophysical impact of the measurements we implemented the new data into a steady state model of the composition of the outer accreted crust of a neutron star. The model has been described in detail in \cite{Lau2018,Schatz2022}. Briefly,
the model uses a full reaction network including electron capture, $\beta$ decay, neutron capture, neutron emission, fusion reactions, and neutron transfer reactions to determine the steady state composition as a function of depth at a constant mass accretion rate. The mass accretion rate is 0.3$\dot{m}_{\rm Edd}$, with $\dot{m}_{\rm Edd}=\EE{8.8}{4}$~g/cm/s being the local Eddington accretion rate in the rest-frame at the neutron star surface. The model provides nuclear energy deposition and cooling rates as a function of depth driven by ongoing accretion once a full accreted crust has formed. The crust temperature profile is kept constant at a fiducial temperature of 0.5~GK. This temperature is in the upper range of expected temperatures and provides a good starting point to identify important Urca cooling processes, which strongly depend on temperature as $\propto  T^5$ \cite{Schatz2013}. This approach enables us to identify the critical reactions that limit crust heating in sources with strong heating mechanisms without being dependent on assumptions for a particular system. The initial composition depends on the surface burning of accreted hydrogen and/or helium. Here we are interested in crust cooling in systems that experience superbursts driven by explosive carbon burning. We therefore choose an initial composition from 1D hydrodynamical superburst models \cite{Keek2012}, see \cite{Jain2025} for details.

The default reaction network uses allowed EC and $\beta$-decay transition strengths from parent ground states to ground and excited states predicted with QRPA-fY. We updated our reaction network with experimental information on key Urca cooling pairs for superburst ashes compositions where available. In particular, we use experimental gs-gs transition strengths $\log ft_{\beta}$ for the $\beta$ decays of $^{29}$Na \cite{baumannGamowTellerBetaDecay1987}, $^{57}$V \cite{mantica$ensuremathbeta$DecayStudies2003}, $^{61}$V \cite{ongDecay61VIts2020},  $^{61}$Cr \cite{crawfordLowenergyStructure$^61mathrmMn$2009}, and $^{59}$Mn \cite{oinonenGroundstateSpin59Mn2001}. The corresponding EC gs-gs transition strengths $\log ft_{EC}$ were calculated as $ft_{EC} = J_{EC}/J_{\beta} \times ft_{\beta}$. For $^{61}$Mn, an older experiment \cite{runteDecayStudiesNeutronrich1985} indicated a gs-gs $\log ft = 4.4$ in agreement with QRPA-fY predictions. However, a more recent measurement indicates $\log ft > 5$ \cite{radulovDecay61Mn2013}. We therefore adopted $\log ft = 5$. For $^{55}$Sc, parent and daughter ground-state spins are inferred to be 7/2 and 1/2, respectively and no gs-gs feeding has been identified in decay experiments \cite{crawfordDecayIsomericProperties2010b}. We therefore remove the gs-gs transitions in $^{55}$Ti$ \leftrightarrow ^{55}$Sc from the network. 

\begin{figure}[tbp]
\includegraphics[width=0.95\linewidth]{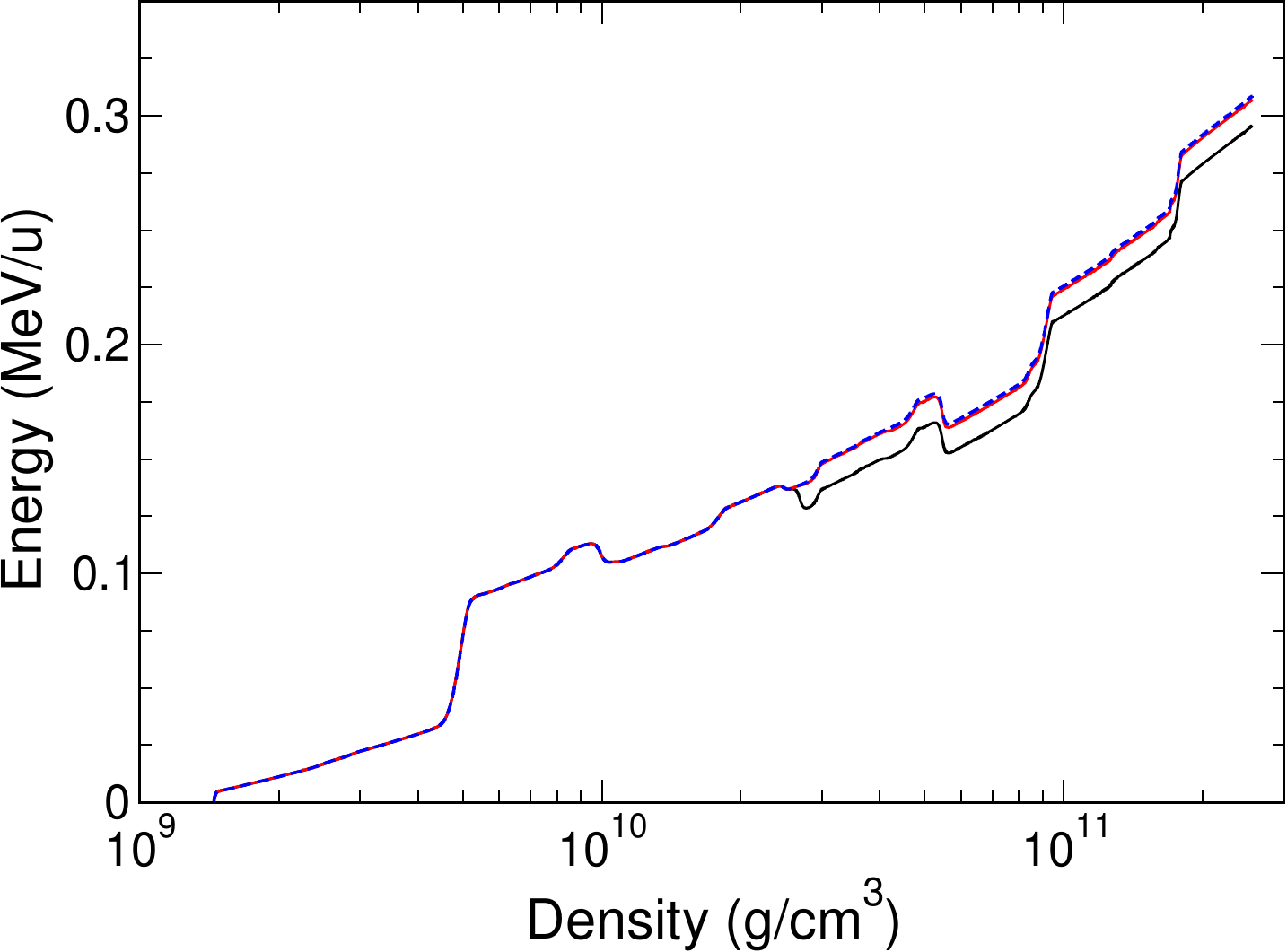}
\caption{Integrated energy produced in the crust as a function of density using nuclear data available prior to this work (black, solid), using in addition the lower limits of our new $\log ft$-values (red, solid) and using the upper limits of our new $\log ft$-values (blue, dashed). Positive jumps indicate a heat source, drops indicate an Urca cooling pair.  
\label{fig:crust}}
\end{figure}

We then performed three model calculations with different implementations for $\beta$ decays and EC for Urca pairs with $\beta$-decay parents $^{57}$Sc, $^{57}$Ti, and $^{59}$Ti: (1) with QRPA-fY predictions, which for $^{57}$Ti agree with the previous high-resolution $\gamma$-spectroscopy measurement \cite{liddickVdecayOddA57Ti2005};(2) with the lower 1$\sigma$ limits of the gs-gs $\log ft$ values obtained in this work; and (3) with the upper 1$\sigma$ limits of the gs-gs $\log ft$ values obtained in this work. Low-lying excited states can also contribute to the Urca cooling rates. We therefore also included our $\beta$-decay transitions to the 270, 175, and 115~keV excited states in the decays of $^{57}$Sc, $^{57}$Ti, and $^{59}$Ti, respectively. The resulting integrated nuclear energy is shown in Fig.~\ref{fig:crust} and the reaction flows of the top Urca pairs are listed in Tab.~\ref{tab:table1}.

In contrast to the theoretical results obtained previously (see for example Fig.~14 in \cite{Lau2018}), cooling rates indicated by drops in the integrated energy are much smaller than the heating rates at 0.5~GK when employing experimental data. Prior to our measurement, the $^{57}$V$\leftrightarrow ^{57}$Ti urca pair was the strongest cooling agent, with $^{29}$Mg$\leftrightarrow ^{29}$Na a strong second. With our new data,   $^{57}$V$\leftrightarrow ^{57}$Ti is weaker by at least an order of magnitude, making now $^{29}$Mg$\leftrightarrow ^{29}$Na and 
$^{55}$V$\leftrightarrow ^{55}$Ti the strongest cooling pairs. The $^{29}$Mg$\leftrightarrow ^{29}$Na gs-gs transition strength is based on  experimental high-resolution $\beta$-delayed $\gamma$-ray spectroscopy data for the ground state feeding \cite{baumannGamowTellerBetaDecay1987}. For $^{55}$V$\leftrightarrow ^{55}$Ti we use theoretical transitions from QRPA-fY. Predicted $\log ft$ values vary between gs and low-lying excited states at 10s of keV from 4.5 to 5.1. These are likely overestimated, given a lower limit from high-resolution $\beta$-delayed $\gamma$-spectroscopy of $\log ft >$5.4 \cite{mantica$ensuremathbeta$decayProperties$^5556mathrmTi$2003}. A measurement with the TAS technique of the $\beta$ decay of $^{55}$Ti would be useful to determine a more stringent limit and potentially rule out this Urca pair.

$^{57}$Ti$\leftrightarrow ^{57}$Sc was negligible before, a factor of 80 weaker than the strongest pair. Our measurements confirmed that this pair is not important, reducing its strength further by another factor of 5--10. For the $^{59}$V$\leftrightarrow ^{59}$Ti pair previous spin assignments indicate the potential for a strong gs-gs transition. Our measurement indicates that this is not the case, and the pair remains unimportant. In fact, the comparison with the QRPA-fY predicted strength function rather supports predictions from recent shell-model calculations \cite{wimmerFirstSpectroscopy61Ti2019} for a 1/2$^-$ $^{59}$Ti ground state that would prevent allowed gs-gs transitions (Fig.~\ref{fig:strength_ti59}). Fig.~\ref{fig:crust} also shows that the remaining uncertainties for Urca cooling via $^{57}$Ti$\leftrightarrow ^{57}$Sc, $^{57}$V$\leftrightarrow ^{57}$Ti, $^{59}$V$\leftrightarrow ^{59}$Ti, and $^{57}$V$\leftrightarrow ^{57}$Ti are unimportant, and that our data provide sufficient constraints for astrophysical applications. 

\begin{table}[ht]
\caption{\label{tab:table1}
The most intense Urca pairs with $\beta$-decay parent charge ($Z$) and mass ($A$) number, after implementing our results, 
that have reaction flows within 2\% of the strongest pair. The ranges indicates the uncertainty from the new logft values determined in this work}
\begin{ruledtabular}
\begin{tabular}{ccl}
   Z  &   A  & $\beta$-decay flow (mol/g)\\ \hline
   22 & 57   & \EE{2.0}{-3}- \EE{6.0}{-3}\\
   22 &  59  & \EE{2.6}{-3}- \EE{6.8}{-3}\\ 
   24 &  59  & \EE{7.4}{-3}\\
   20 &  55  & \EE{7.6}{-3}\\
   23 &  57  & \EE{1.0}{-2}\\
   23 &  55  & \EE{1.6}{-2}\\
   23 &  59  & \EE{2.0}{-2}\\
   22 &  55  & \EE{6.5}{-2}\\
   11 &  29  & \EE{8.3}{-2}\\
\end{tabular}
\end{ruledtabular}
\end{table}

\section{Conclusions}
We performed measurements of the feeding of final states in the $\beta$ decays of $^{57}$Sc, $^{57}$Ti, and $^{59}$Ti using total absorption $\gamma$-spectroscopy and measurements of $\beta$-delayed neutron emission. Our data support the recently proposed revised 1/2$^-$ ground-state spin assignments of neutron-rich $^{57}$Ti and $^{59}$Ti isotopes. For the decay of $^{57}$Sc we quantitatively confirm the predominant feeding of the first excited state in $^{57}$Ti that can be explained well with a spin 1/2$^-$ $^{57}$Ti ground state. For the decay of $^{57}$Ti we find a much smaller feeding of the $^{57}$V ground state of 3(2)\% compared to 54\% from previous work with high-resolution spectroscopy. From the decay of  $^{59}$Ti we present the first low-lying level scheme of $^{59}$V and again find a relatively small branch to the ground state of only 8(2)\%. 

The relatively small gs-gs branchings found in this work agree with the low strength of the $^{59}$V$\leftrightarrow ^{59}$Ti Urca cooling pair in neutron stars predicted by the QRPA-fY theory, but reduce the strength of the $^{57}$V$\leftrightarrow ^{57}$Ti pair significantly compared to theory predictions. With this work, when taken together with previous investigations, a trend is emerging that predictions of Urca cooling on accreted neutron star crusts are significantly reduced when employing experimental data, especially when employing the more reliable total absorption spectroscopy technique. 


The only remaining currently predicted potentially relevant Urca pair without experimental constraint for accreted crusts formed from the ashes of superbursts is $^{59}$Cr$\leftrightarrow ^{59}$V. A measurement of the ground-state branch of the $\beta$ decay of $^{59}$V would therefore be desirable. The strength of the two strongest Urca pairs, $^{29}$Mg$\leftrightarrow ^{29}$Na and $^{55}$V$\leftrightarrow ^{55}$Ti, still rely on data from high-resolution $\gamma$-spectroscopy that may suffer from significantly underestimated uncertainties. 

\begin{acknowledgments}
This work has been supported by NSF grants PHY-2514797, PHY-2209429, PHY-1913554, PHY-1102511. The work strongly benefited from activities supported through NSF awards PHY 14-30152 (Joint Institute for Nuclear Astrophysics), OISE-1927130 (International Research Network for Nuclear Astrophysics), and DOE award DE-SC0023128 (CeNAM).

\end{acknowledgments}

\bibliography{main,Kirby2}

\end{document}